# Providing an Object Allocation Algorithm in Distributed Databases Using Efficient Factors

**Arash Ghorbannia Delavar[1], Golnoosh keshani[2]**

**[1] Department of Computer Engineering and Information Technology, Payam Noor University,**
**PO BOX 19395-3697, Tehran, IRAN**
***a_ghorbannia@pnu.ac.ir***

**[2] Department of Computer Engineering and Information Technology, Payam Noor University,**
**PO BOX 19395-3697, Tehran, IRAN**
***golnoosh_keshani@yahoo.com***

## Abstract

Data replication is a common method used to improve the performance of data access in distributed database systems. In this paper, we present an object replication algorithm in distributed database systems (*ORAD*). We optimize the created replicated data in distributed database systems by using activity functions of previous algorithms, changing them with new technical ways and applying *ORAD* algorithm for making decisions. We propose *ORAD* algorithm with using effective factors and observe its results in several valid situations. Our objective is to propose an optimum method that replies read and write requests with less cost in distributed database systems. Finally, we implement *ORAD* and *ADRW* algorithms in a PC based network system and demonstrate that *ORAD* algorithm is superior to *ADRW* algorithm in the field of average request servicing cost.

***Keywords:*** object *replication, Database system, Servicing cost, ADRW algorithm, ORAD algorithm.*

## 1. Introduction

We are presently moving towards a distributed, wholly interconnected information environment. Generally, in distributed database systems an object will be accessed, i.e. read and written, from multiple processors [4]. The requests for an object that come from a processor may be answered in two ways and the first is when the system has the object on its local memory and the requests are responded locally and the second is when the system does not have the object on its local memory and must send the request to another system that has it on its local memory and can send (should be a server) it to the requesting system. Replication strategies are part of most distributed storage mechanisms [10].

Replication reduces data access time and improves the performance of the system [2]. One thing that is important in distributed databases is to warrant the consistency of multiple replicas of an object in multiple systems. So every change to an object must be transferred to all the other available replicas, this will incur considerable communication cost [1].

Generally, when more copies of an object are created, the average write request servicing cost will increase, but the average read request servicing cost will decrease. Therefore, in order to manage the number of copies of objects, we need an efficient replication mechanism that can be optimized to respond to read and write requests with minimal cost in distributed database systems. A replication mechanism specifies which file should be replicated, when to create new replicas and where the new replicas should be placed [5].

In this paper, we introduce *ORAD* algorithm with a cost model and a correct mechanism in designing request windows. As the distributed database systems are dynamic, there is not any information about the number of requests. Thus the decisions at each stage of *ORAD* algorithm are based on the history of recent requests. Then we implement *ADRW* and *ORAD* algorithms and analyze the performance of both algorithms in several valid situations.

## 2. Related work

Various static and adaptive data replication algorithms and on-line problems in distributed systems were proposed [8], [10], [11], [12]. One of them is *SA* algorithm [6] (static algorithm).

### 2.1 *SA* algorithm

The allocation scheme of a distributed system determines how many replicas of each object are created and to which processors these replicas are allocated [6]. At all times, *SA* keeps a fixed allocation scheme *Q* which is of size *t*. All the processors in the system know which are the processors of *Q*. *SA* performs read-one-write-all. Namely, in response to a write request issued by a processor *p*, *SA* sends the object from p to each one of the processors in Q. In turn, each processor of *Q* outputs the object in its local database. In response to a read request issued by a processor *p*, *SA* requests a copy of the object from some processor $y \in Q$; in turn, *y* retrieves the replica from its local database and sends it to *p* [6].

Another algorithm that was introduced after *SA* algorithm is *DA* algorithm (Dynamic algorithm)

## 2.3 *DA* algorithm

The *DA* algorithm receives as parameters a set *F* of *t* − 1 processors, and a processor *p* that is not in *F*. The processors of *F* are called the servers, and *p* is called the floating processor.

All the processors in the system know the id of the processors in $F \cup \{p\}$. The initial allocation scheme consists of $F \cup \{p\}$. Subsequently, at any point in time all the servers are in the allocation scheme and at least one additional processor is there as well; however, the floating processor is not necessarily in the allocation scheme. For example, for non server, non floating processors *q* and *r*, $F \cup \{q\} \cup \{r\}$ is a possible allocation scheme at some point in time [6].

The *DA* algorithm services read and write requests as follows. A read request from a processor of the allocation scheme is satisfied by inputting the object from the local database. A read request from a processor *r* outside the allocation scheme is satisfied by requesting a copy of the object from some server processor *u*; *r* saves the object in its local database (thus joining the allocation scheme), and u remembers that r is in the current allocation scheme by entering r in *u*'s "join-list." The join-list of *u* consists of the set of processors that have read the object from *u* since the latest write.

A write request from some processor *q* outputs the object to the local database at *q* and sends it to all the servers; then, each server outputs the object in its local database.

If *q* is a server, then *q* also sends a copy of the object to the floating processor (in order to satisfy the availability constraint). Additionally, the write request results in the invalidation of the copies of the object at all the other processors (since their version is obsolete). This is done as follows.

Each server, upon receiving the write, sends an "invalidate" control-message to the processors in its "join-list" (except that, obviously, if *q* is in some join list, the invalidation message is not sent to *q*). To summarize the effects of a write, consider the allocation scheme *A* immediately after a write from a processor *q*. If *q* is in *F*, then $A = F \cup \{p\}$, and if *q* is not in *F*, then $A = F \cup \{q\}$ [6].

## 2.3 *ADRW* Algorithm

The goal of the *ADRW* algorithm is to dynamically adjust the replication and allocation of objects in order to minimize the total servicing cost of the requests coming to the distributed database system [1, 3]. The servicing cost is defined to consist of three components as follows;

$C_c$: Cost of sending the query for the object.

$C_{io}$: Cost of fetching/updating the object to/from the local memory of the processor.

$C_d$: Cost of transferring the object from the main memory of the hosting (i.e. data) processor to the requesting (i.e. non-data) processor.

S(*o*): Initial allocation servers for object *o*.

The processor is considered a data processor for a particular object if the object is hosted in the local memory of the processor. All other processors are non-data processors for the object. Assuming we have three processors $p_1$, $p_2$, and $p_3$ and $p_2$ is the data processor for object *o*. The cost for $p_2$ to access object *o* is one unit of time. Moreover, $p_2$ will create a *k*-bit size window corresponding to object *o*. For every new request coming to $p_2$ for object *o* from $p_1$, a 0 is added to Win(*o*, $p_1$), while a 1 is added to Win(*o*, $p_3$) for every new request coming to $p_2$ from $p_3$ for object *o*. If another process, say $p_3$, is writing to the object *o*, then $p_2$ will add 1 to the window. So, if the number of read, $N_r$, from $p_1$ is greater than the write, $N_w$, from $p_3$, then $p_2$ will make a replication for *o* to $p_1$ with its window and add $p_1$ to the data_list(*o*) which is a list of all the processors that have a replica of the object *o*. $p_1$ now is a data processor. It will save the object in its local memory and access it directly. If any write to the object arrived to $p_2$ then it will update the object and send the update to all the processors that hold the object found in the data_list(*o*). Now, if processor $p_1$ reads the object, it will add 0 to the window and 1 if others write to it. If the number of writes is greater than the number of reads, then it will delete the replication and return the window to the owner processor $p_2$ [3].

## 3. Proposed Algorithm

In this approach, we suggest a dynamic replication algorithm method. A replication method is a way of describing the actual replication process. For the implementation of the method of *ORAD* algorithm, we change the method of *ADRW* algorithm and discuss other cost factors in addition to the cost of the three factors mentioned in *ADRW* algorithm. The algorithm changes the replication scheme, i.e., number of replicas and their location in the distributed database system, to optimize the amount of communication [1]. We also introduce flag bits in servers and say how they are created and initialized.

We consider a distributed database system with n nodes (n processors), denoted as $p_1$, $p_2$ ,..., $p_n$. Each node has a processor and a local memory. All the local memories are private and accessible only by their local processors and assume that there exist at least $1 \leq t \leq n$ replicas in the system.

In *ORAD* algorithms, we divide the processors into two parts based on their recent access history, denoted as data processors and non-data processors. To illustrate the operation of *ORAD* algorithm, at first we assume that all processors are non-data processor for all objects and do not have the objects on their local memories. Each server creates a flag bit $F^{Pi}_o$ for object $o$ and processor $p_i$ on its local memory if processor $p_i$ sends at least one request for object $o$ to it. While processor $p_i$ is selected by *ORAD* algorithm for object $o$ as a data processor, $p_i$ saves the object on its local memory. Furthermore, while a data processor $p_i$ is changed to the non-data processor for object $o$ by the algorithm, the server updates the flag bit to 1. The non-data-processor $p_i$ keeps object $o$ temporary until the server sends the invalid message to it. After receiving the invalid message, the non-data processor deletes object $o$ from its local memory. We dissect the method of *ORAD* algorithm with an example.

In Fig. 1, at first we consider all processors ($p_1$, $p_2$, $p_3$, $p_4$, $p_5$, $p_6$) as a non-data processor and $s_1$, $s_2$ as a server for an object $o$.

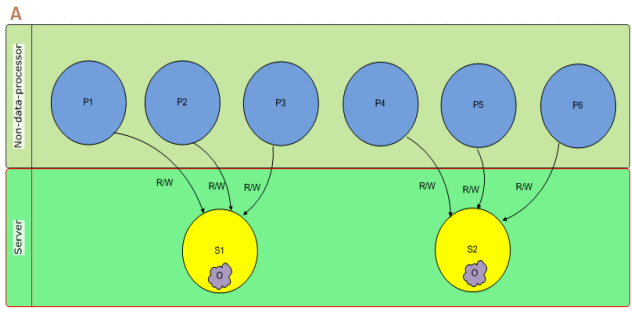

Fig. 1 Step 1 of the method of *ORAD* algorithm

After receiving these requests "$W^{P1}$, $R^{P4}$, $R^{P4}$, $W^{P6}$, $R^{P2}$, $R^{P5}$, $W^{P1}$" for object $o$, the role of processors is changed as shown in Fig. 2. The processors $p_2$, $p_4$ and $p_6$ are changed to the data processor by the algorithm and they keep the copy of object $o$ in their local memory ($o'$).

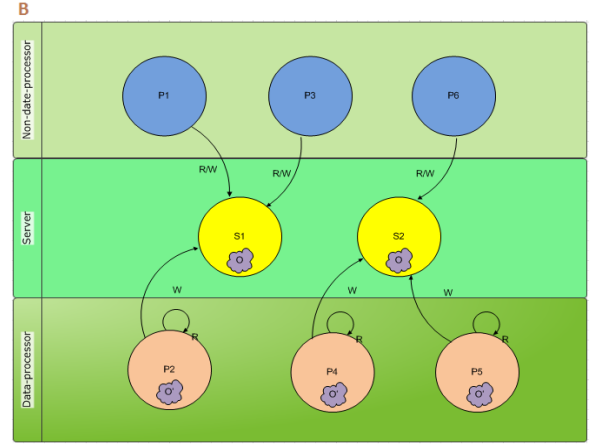

Fig. 2 Step 2 of the method of *ORAD* algorithm

In addition to the requests as mention above, these requests are also received; "$R^{P5}$, $R^{P4}$, $R^{P2}$, $W^{P5}$, $R^{P5}$, $W^{P5}$, $W^{P3}$, $R^{P4}$, $W^{P5}$". Now the request sequence is "$W^{P1}$, $R^{P4}$, $R^{P4}$, $W^{P6}$, $R^{P2}$, $R^{P5}$, $W^{P1}$, $R^{P5}$, $R^{P4}$, $R^{P2}$, $W^{P5}$, $R^{P4}$, $R^{P5}$, $W^{P5}$", so *ORAD* algorithm decides to remove $p_5$ from data-list($o$), but $p_5$ keeps $o'$ in its local memory temporary until receiving a write request on the object $o$ such as $W^{P3}$ (Fig. 3).

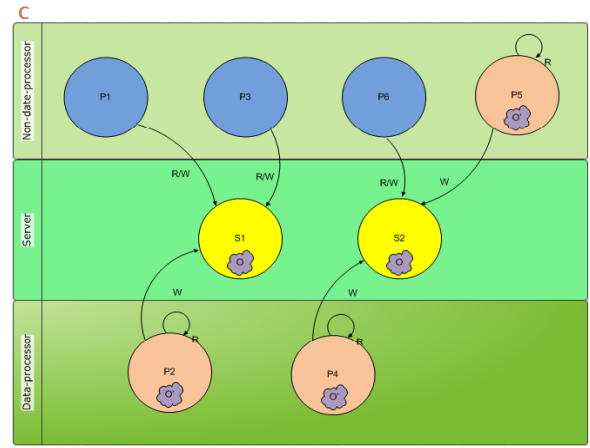

Fig. 3 Step 3 of the method of *ORAD* algorithm

In the end $p_5$ deletes $o'$ from its local memory (Fig. 4, Fig. 5).

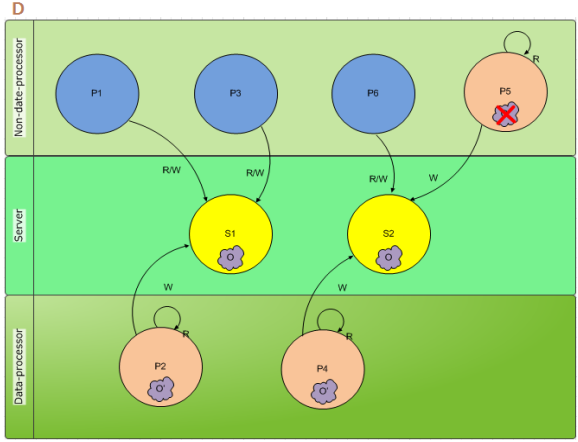

**Fig. 4** Step 4 of the method of *ORAD* algorithm

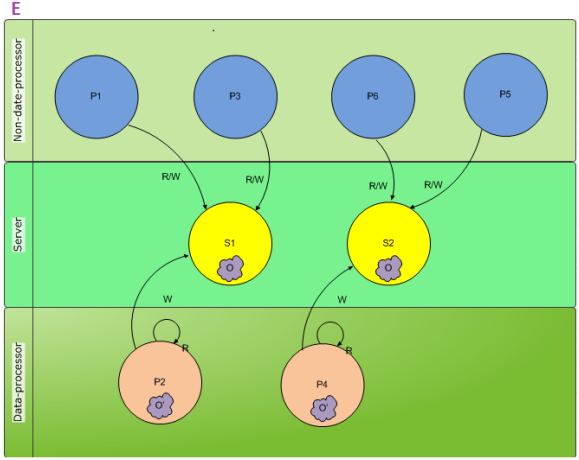

**Fig. 5** The last step of the method of *ORAD* algorithm

Table 1 presents a glossary of notation used throughout this paper.

TABLE 1: GLOSSARY OF NOTATION

| | |
|---|---|
| *Req* | Request |
| $R^{Pi}_o$ | Read request from processor pi for object *o* |
| $W^{Pi}_o$ | Write request from processor Pi for object *o* |
| $Cost_A(Req)$ | Cost of servicing a request *Req* using an algorithm *A* |
| $S(o)$ | Server set of an object *o* |
| *Ao* | Allocation scheme of object *o* |
| $MR_w(o,p_i)$ | Message & request window |
| $R_{Ld}$ | local read request from a data processor for an object |
| $R_{Rn}$ | Remote read request from a non-data processor for an object |
| $R_{Ln}$ | local read request from non-data processor |
| $W_{Ld}$ | local write request from a data processor for an object |
| $W_{Rd}$ | Remote write request that is propagated form a server for an object |
| *Inv* | Invalid control message from a server for an object |
| *F* | Flag bit |
| $F^{Pi}_o$ | Flag bit of processor pi for object o |

### 3.3 Cost model

We now present our method to compute the cost of servicing a read or write request.

Read request: consider servicing a read request ($R^{Pi}_o$) from $p_i$ for object *o* and let $A_o$ be the allocation scheme of object *o* on this request and *F* be the flag bit in server $p_j$ (the nearest server in S(*o*) to $p_i$) for object *o* and processor $p_i$ . Then,

$$\text{Cost}_{\text{ORAD}}(R^{Pi}_o) = \begin{cases} 1 & \text{if } p_i \in A_o \\ 1 & \text{if } p_i \notin A_o \text{ and } F \text{ is } 0 \\ 1+C_c+C_d & \text{if } p_i \notin A_o \text{ and } F \text{ is } 1 \text{ and } R \text{ is not saving request} \\ 2+C_c+C_d & \text{if } p_i \notin A_o \text{ and } F \text{ is } 1 \text{ and } R \text{ is saving request} \end{cases}$$

(1)

In Eq. (1), While $p_i \in A_o$, it means that $p_i$ is a data processor for object *o*. Thus for every read request, it is enough to read object *o* from its local memory, incurring only *I/O* cost. We assume that $C_{io}$=1 (like *ADRW* algorithm). On the other hand, if $p_i \notin A_o$ and *F* is 0, then $p_i$ is a non-data processor for object *o*, but the object is still on its local memory (the object is still valid), incurring only $C_{io}$ cost and if $p_i \notin A_o$ and *F* is 1, it means that $p_i$ is a non-data processor for object *o* and does not have the object on its local memory. So $p_i$ will send a read request to the nearest server (since the server set is known to each processor), say $p_j$, in S(*o*), incurring $C_c$ units of cost. After receiving the read request, $p_j$ will then retrieve object *o* from its local memory and send it to $p_i$, incurring ($C_{io}$ + $C_d$) units of cost. Finally if $p_i$ saves object *o* into its local memory (saving-read), then the servicing cost will be one unit higher than if $p_i$ does not save object *o* into local memory (non-saving- read). As *ADRW* algorithm [1], Once server $p_j$ decides that the request is a saving-read request, $p_j$ will add processor $p_i$ into a data processor list, denoted by data-list(*o*) (since $p_i$ now is a data processor), so that following write requests for the object *o* can be propagated to the processors in data-list(*o*) for data consistency.

Write request: Consider servicing a write request ($W^{Pi}_o$) from processor $p_i$ for object *o*. Let $A_o$ be the allocation scheme of object *o* on the request before servicing this request, $A'_o$ be the allocation scheme of object *o* after servicing this request, $N_{Fo}$ be the number of flag bits with value 1 for object $o$ before servicing this request and $N'_{Fo}$ be the number of flag bits with value 1 for object $o$ after servicing this request.

$$\text{Cost }_{ORAD}(W^{Pi}_o) = \begin{cases} (|A_o| - 1)\ C_d + |A'_o|\ + N_{Fo} + N_{Fo}\ C_c + N'_{Fo} \\ \qquad\qquad\qquad\qquad\qquad \text{if } p_i \in A_o \\ |A_o|\ C_d + |A'_o|\ + N_{Fo} + N_{Fo}\ C_c + N'_{Fo} \\ \qquad\qquad\qquad\qquad\qquad \text{otherwise} \end{cases}$$
(2)

In order to maintain the object consistency, when a write request for object $o$ is issued, the new version of object $o$ should be transferred to all data processors. Each data transfer will incur $C_d$ units of cost. If $p_i \in A_o$, then object $o$ will be transferred to all the data processors in $A_o$ other than $p_i$, incurring $(|A_o| - 1)\ C_d$ units of cost. Otherwise, object $o$ will be transferred to all the data processors in $A_o$, incurring $(|A_o|\ C_d)$ units of cost [1]. The processor $p_i$ first sends the new version to all the servers in S($o$). All the servers then propagate the new version to the processors in their respective data-list($o$) to maintain the object consistency. According to our *ORAD* algorithm, some data processors in $A_o$ which are not in S($o$), may exit the allocation scheme to minimize the total servicing cost of future requests. Only those processors in $A'_o$ save the new version into their respective local memories, incurring $|A'_o|$ units of *I/O* cost [1]. Furthermore, for each flag bit of object $o$ which is 1 in the server of object $o$, the server should send an invalid message to the non-data processor corresponding to that flag bit, incurring $(N_{Fo}\ C_c)$ units of cost and then the server updates all flag bits of object $o$ to 0 , incurring $((N_{Fo}\ C_{io}) = (N_{Fo}))$ units of cost. Finally, after servicing the write request, according to new allocation scheme $A'_o$, some data processor may be changed to the non-data processor for object $o$. Therefore the flag bits of object $o$ should be 0 by the server, incurring $((N'_{Fo}\ C_{io}) = (N'_{Fo}))$ units of cost.

## 3.2 Distributed message & request window mechanism

As mentioned above, a non-data processor $p_i$ refers to its nearest server $p_j$ for servicing its requests on object $o$. Then server $p_j$ creates a message & request window $MR_w(o,p_i)$ for processor $p_i$ unless $MR_w(o,p_i)$ already exists. For every message or request related to object $o$ that $p_j$ receives from $p_i$, $p_j$ initializes $MR_w(o,p_i)$. When *ORAD* algorithm decides to select $p_i$ as a data processor, server $p_j$ sends $MR_w(o,p_i)$ to $p_i$ for saving other requests and messages because now, $p_i$ is a data processor for object $o$ and all messages and requests should be sent to it. Furthermore, when *ORAD* algorithm decides to remove $p_i$ from data-list($o$), the server sets $F^{Pi}_o = 1$, but $p_i$ will not transfer $MR_w(o,p_i)$ to the server until the server sends the invalid message $Inv_o$ to it[1].

## 3.3 Read request:

Servicing a read request on object $o$ which is issued by a non-data processor $p_i$, is done in two ways;

- If the non-data processor has object $o$ on its local memory ($F^{Pi}_o$ is 1), it means that $p_i$ have already been a data processor for object $o$ and ORAD algorithm removed it from data-list($o$), but $p_j$ have had temporarily object $o$ on its local memory yet. In this case $p_i$ has not transferred object $o$ to the server yet. So $p_i$ services the request locally and then inserts $R_{Ln}$ in $MR_w(o,pi)$.

- If the non-data processor $p_i$ does not have the object, it should refer to the server. After servicing the request, Because the server has $MR_w(o,p_i)$, inserts $R_{Rn}$ in $MR_w(o,p_i)$.

## 3.4 Write request:

When a processor $p_i$ wants to write on an object $o$, at first sends the write request to the server. After that the server sends the new version of object $o$ to all data processors. Note that if $p_i$ is a data processor, it is not required that the server sends the new version of object $o$ to $p_i$ because $p_i$ has the new version of object $o$. So if a data processor receives a write request for object $o$, at first the processor updates the object on its local memory and after that if the request comes from itself, it inserts $W_{Ld}$ in $MR_w(o,p_i)$ and if it is propagated form the server, it inserts $W_{Rd}$ in $MR_w(o,p_i)$.

## 3.5 Invalid message:

This message is sent from server $p_i$ to non-data processor $p_j$ that has object $o$ temporarily. The message shows that a new version of object $o$ is created and object $o$ in $p_i$ is invalid. When $p_i$ receives the invalid message, inserts $I_{nv}$ in $MR_w(o,p_i)$, removes object $o$ on its local memory and transfers $MR_w(o,p_i)$ to $p_j$ for saving its future requests.

## 3.6 Flag bit ($F^{Pi}_o$)

As mentioned above, server $p_j$ creates a bit flag ($F^{Pi}_o$) for each non-data processor $p_i$ sends at least one read or write

---

[1] It means that the object o on its local memory is invalid and changed by another processor

request to it for object $o$. When a data processor $p_k$ is changed to the non-data processor for an object $o$ by *ORAD* algorithm, server $p_j$ will insert 1 to flag bit $F^{Pk}_o$ and after that if object $o$ will be changed, server $p_j$ will send invalid message $Inv^{Pk}_o$ to $p_k$ and update value of the flag bit to 0. So can conclude that the number of changing flag bit $F^{Pk}_o$ is twice the number of invalid message $Inv^{Pk}_o$.

### 3.7 Calculating the servicing cost of the requests:

Now, we want to compute cost of mentioned requests and messages in message & request window.

**$R_{ld}$** : it is a local read request that is issued by a data processor for an object and will be serviced locally by reading from the local memory of data processor $p_i$, incurring $C_{io}$ units of cost (1).

**$W_{ld}$** : it is a local write request which is sent to data processor $p_i$ for object $o$ by its self, incurring $C_{io}$ units of cost for updating the new value of object $o$. It not required that the server sends the new version of object $o$ to $p_i$ because $p_i$ has changed the object itself and it has the new version of object $o$.

**$W_{Rd}$** : it is a remote write request that is propagated form server $p_j$ to data processor $p_i$ for object $o$, incurring $(C_d+1)$ units of cost, $C_d$ units of cost for sending data message and one unit of cost for updating the object on the local memory of data processor $p_j$.

Although it was said in *ADRW* algorithm that in the same write request $W_{ld}$, no need to send the new version of object $o$, but when it calculated the servicing cost, it defined only one type of write request, incurring $(C_d + 1)$ units of cost [1] and it is one of the main differences between *ORAD* algorithm and *ADRW* algorithm.

**$R_{Rn}$** : It is a remote read request that is sent from non-data processor $p_k$ to server $p_j$ for an object $o$, incurring $(C_d + C_c + 1)$ units of cost; $C_c$ units of cost for of sending the query from non-data processor $p_k$ to the server for the object, 1 unit of cost for fetching the object from the local memory of the server and $C_d$ units of cost for transferring the object from the local memory of the server to the non-data processor.

**$R_{Ln}$** : It is a local read request from non-data processor $p_k$ that will be serviced locally by reading from its local memory, incurring 1 unit of cost. Because non-data processor $p_k$ does not have the object on its local memory, or has it temporary, no write request will be reached to it.

**$Inv$** : It is an invalid control message that is sent from a server to a non-data processor, incurring $C_c$ units of cost. The non-data processor has an object on its local memory temporary. When the server sends the invalid message to the non-data processor, it means that object has been changed and the copy of object that is placed on the local memory of the non-data processor is invalid and should be removed from the local memory of the non-data processor.

### 3.8 Updating the flag bit:

This operation is performed in two modes by a server. The first is when a data processor is changed to the non-data processor by *ORAD* algorithm and the server updates the flag bit to 1, incurring $C_{io}$ units of cost and the second is when the server sends invalid message $Inv$ to the non-data processor and the server updates flag bit $F^{Pi}_o$ to $0$, incurring $C_{io}$ units of cost. So the cost of all the operations is twice the cost of all invalid messages $Inv$ for an object $o$, incurring $N_{Inv}^o \times 2 \times C_{io}$ units of cost ($2 \, N_{Inv}^o$).

Due to the cost calculated in above, *ORAD* algorithm decides to select a processor $p_i$ as a data processor or non-data processor for an object $o$ by comparing the cost in each state.

Cost of being a data processor;
$$N_{Rld} + N_{Wld} + N_{Wrd} \, (C_d + 1) \hspace{2cm} (3)$$

Cost of being non-data processor;
$$N_{Rln} + N_{Rrn} \, (C_d + C_c + 1) + N_{Inv} \, C_c + 2 \, N_{Inv}. \hspace{1cm} (4)$$

Consider $N_{Tr}$ as total number of read requests and $N_{Tw}$ as total number of write requests, so;
$$N_{Tr} = N_{Rld} \quad , N_{Tw} = N_{Wld} + N_{Wrd} \quad \text{if } p_i \in A_o \hspace{1cm} (5)$$
$$N_{Tr} = N_{Rln} + N_{Rrn} \quad , N_{Tw} \quad \quad \text{if } p_i \notin A_o \hspace{1cm} (6)$$

Thus;
Cost of being data processor;
$$N_{Tr} + N_{Wld} + (N_{Tw} - N_{Wld}) \, (C_d +1) = N_{Tr} + N_{Tw} \, C_d + N_{Tw} - N_{Wld} \, C_d \hspace{1cm} (7)$$

Cost of being non-data processor;
$$N_{Rln} + (N_{Tr} - N_{Rln}) \, (C_d + C_c + 1) + N_{Inv} \, C_c + 2 \, N_{Inv} = N_{Tr} + N_{Tr} \, (C_c + C_d) - N_{Rln} \, (C_d + C_c) + N_{Inv} \, C_c + 2 \, N_{Inv} \hspace{1cm} (8)$$

If our *ORAD* algorithm found that;
$N_{Tr} + N_{Tw} \, (C_d + 1) - N_{Wld} \, C_d \leq N_{Tr} + N_{Tr} \, (C_c + C_d) - N_{Rln} \, (C_d + C_c) + N_{Inv} \, C_c + 2 \, N_{Inv}$, i.e.,
$$N_{Tw} \, (C_d + 1) - N_{Wld} \, C_d \leq N_{Tr} \, (C_c + C_d) - N_{Rln} \, (C_d + C_c) + N_{Inv} \, (C_c + 2) \hspace{1cm} (9)$$
then, server $p_j$ would consider $p_i$ as a data processor and if the algorithm found that;
$$N_{Tw} \, (C_d + 1) - N_{Wld} \, C_d > N_{Tr} \, (C_c + C_d) - N_{Rln} \, (C_d + C_c) + N_{Inv} \, (C_c + 2) \hspace{1cm} (10)$$
then, removes $p_i$ from data-list($o$).

When *ORAD* algorithm decides to add $p_i$ to data-list($o$), the server transfers $MR_w(o,p_i)$ to $p_i$. So $p_i$ can register next requests or messages. Furthermore, when *ORAD* algorithm decides to remove $p_i$ from data-list($o$), $p_i$ does not transfer $MR_w(o,p_i)$ to the server until removes object $o$ that is temporary on its local memory[1].

We refer to this whole process of *ORAD* algorithm as tree policies. The first is Test-Enter-Data-list (*TED*) policy, the second is Test-Exit-Data-list (*TXD*) policy and the third is Test-Flag (*TF*) policy. The pseudocode in Table 2 presents the *TED* policy of our request & message window mechanism in server $p_j$ and in table 3 presents the *TXD* policy of it in data processor $p_i$ and in Table 4 presents the *TF* policy of it in non-data processor $p_k$.

In Table 2, the Test-Enter-Data-list (*TED*) pseudocode presents the *TED* policy of our request & message window mechanism in server $p_j$ for object $o$. We assume that the arriving request *Req* is issued from $p_i$ for object $o$.

<div align="center">Table 2: Test-Enter-Data-List (TED)</div>

```
If (Req is a read request)  /*R^{Pi}_o*/
    {if (p_i==p_i) /*the read request Req is issued from p_j itself*/
        {No change to the message & request window in
        p_i;} /*satisfy Req locally*/
    Else /*p_i ≠ p_j , p_i must be a non-data processor*/
        {if Req is the first read request
            {Generate an initial MR_w(o,p_i);}
        Insert R_Rn into MR_w(o,p_i) ;
        Send o to p_i;
        If (N_Tw (C_d + 1) - N_Wld C_d ≤ N_Tr (C_c + C_d) - N_Rln (C_d +
        C_c) + N_inv (C_c + 2))
            {Indicate p_i to enter Ao;
            Add processor p_i into data-list(o);
            Transfer MR_w(o,p_i) to p_i;
            Delete MR_w(o,p_i) in p_j;
            p_i saves object o; /*data processor*/
            }
        }
    }
Else /*Req is a write request for object o*/
    {Send invalid control message to all non-data processors
    p_K that F^{Pki}_o in p_j equals 1;
    Write on object o;
    Update all flag for object o in p_j to 0;
    Insert W_Rd into all existing message & request windows
    for object o in p_j except MR_w(o,p_i) if it exists;
    Send the new version of object o to all data processors of
    object o;
    }
```

In Table 3, there is another operation called Test-Exit-Data-list (*TXD*) policy which is processed in a data processor but not in a server, of an object $o$. We assume that there is a data processor of an object $o$ $p_i$ ($p_i \notin$ S($o$)), and its nearest server in S($o$) is $p_j$.

<div align="center">Table 3: Test-Exit-Data-list (TXD)</div>

```
If (Req is a read request) /*it must be issued by p_i*/
    {insert R_ld into MR_w(o,p_i);} /*satisfy Req locally*/
Else /*Req is a write request*/
    {if (Req is issued by p_i)
        {Send the write request to p_j;
        Write on object o;
        Insert W_ld into MR_w(o,p_i);}
    Else /*Req is propagated from p_j*/
        {Write on object o;
        Insert W_Rd into MR_w(o,p_i);
        If ( N_Tw (C_d + 1) - N_Wld C_d > N_Tr (C_c + C_d) - N_Rln (C_d
        + C_c) + N_inv (C_c + 2))
            {Indicate p_i to delete p_i from data-list(o) and also to
            update F^{Pi}_o to 1;}
        }
    }
```

In Table 4, the third operation called Test-Flag (*TF*) policy which is processed in a non-data processor of an object $o$ $p_k$, and its nearest server in S($o$) is $p_j$.

<div align="center">Table 4: Test-Flag (TF)</div>

```
If Req is a read request /*Satisfy Req locally*/
    {Insert R_ld into MR_w(o,p_i);}
Else /*Req is an invalid message*/
    {Insert Inv into MR_w(o,p_i) ;
    Transfer win(o,p_k) to p_j ;
    Delete o from the local memory of p_k;
    }
```

## 4. Experimental results

In this section, we implement the *ORAD* algorithm and *ADRW* algorithm in a real-life system and study their performance behavior under a variety of situations. We compare the performance of the *ORAD* algorithm with the *ADRW* algorithm.

We consider a distributed database system with the following assumptions; In addition, we set the *I/O* cost, control message transferring cost and data message transferring cost as $C_{io} = 1$, $C_c = 5$, and $C_d = 10$, respectively, in our experiments.

---

[1] Processor $p_i$ removes object $o$ from its local memory after receiving the invalid message for object $o$.

We introduce;

$o$ = Object
$p$ = Processor
$s$ = Server
$s_i = \{s_1, s_2\}$
$p_i = \{p_1, p_2, p_3, p_4, p_5, p_6, p_7\}$
$o_i = \{o_1, o_2, o_3, o_4, o_5\}$

At first, we consider that all processor are non-data processors for all objects and then we observe the results of these tow algorithms by using many different states of requests.

We show the cost performance algorithms in the following experiments where the maximum size of request is 100 and each node has the same probability of read/write request. Note that the number and type of requests in each state is random. For example in the first row, the number of random requests is 90 and the mean random probability of read requests is 0.1.

TABLE 5: RANDOM REQUEST TABLE

| Maximum size of request | 100 |
|---|---|
| number of request | Mean random probability of read request |
| 90 | 0.1 |
| 34 | 0.2 |
| 99 | 0.3 |
| 48 | 0.4 |
| 87 | 0.5 |
| 22 | 0.6 |
| 67 | 0.7 |
| 75 | 0.8 |
| 43 | 0.9 |

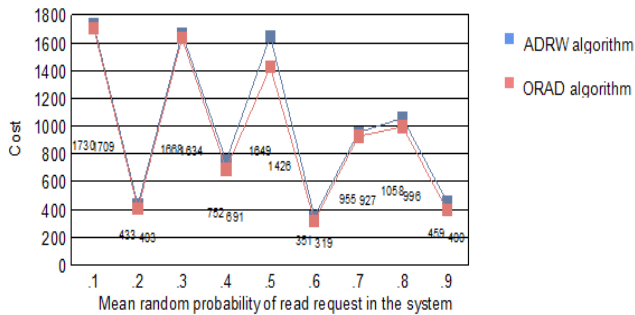

Fig. 6  Cost performance of *ORAD* and *ADRW* algorithm when the maximum number of request is 100 and each node has the same probability of read/write request.

Fig. 6 shows that in random requests, *ORAD* algorithm is more adaptive in terms of the average cost of servicing a request. In all probability of read request, we observe that the *ORAD* algorithm can perform much better than *ADRW* algorithm in random requests.

We can see in Fig. 6 that performance of *ORAD* algorithm improved about 6.07 percent compared with *ADRW* algorithm.

Now, we want to show the cost performance of these tow algorithms where each node has the same probability of read/write request and the number of requests is fixed in all states, but the type of requests (read/write) in each state is random. In distributed database systems, the number of requests is not fixed, but we suppose it to compare the performance of *ORAD* algorithm and *ADRW* algorithm for various read request probabilities in same situations (same number of request). The cost performance of *ORAD* and *ADRW* are shown in Fig. 7 and Fig. 8.

TABLE 6: RANDOM REQUESTS TABLE WITH FIXED NUMBER OF REQUEST (100)

| The number of request=100 | | |
|---|---|---|
| Cost of *ORAD* algorithm | Cost of *ADRW* algorithm | Mean probability of read request |
| 1251 | 1342 | 0.1 |
| 1344 | 1383 | 0.2 |
| 1720 | 1773 | 0.3 |
| 1780 | 1812 | 0.4 |
| 1960 | 1976 | 0.5 |
| 1774 | 1790 | 0.6 |
| 1533 | 1609 | 0.7 |
| 1291 | 1305 | 0.8 |
| 1009 | 1000 | 0.9 |

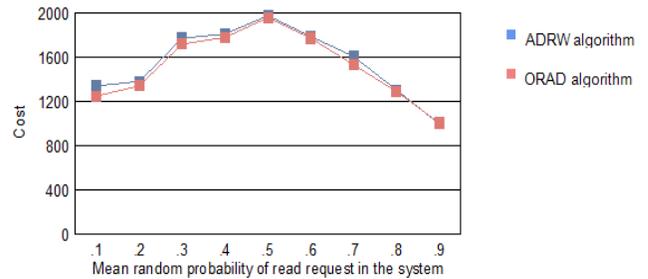

Fig. 7  Cost performance of *ORAD* and *ADRW* algorithm when the number of request is fixed (100) and each node has the same probability of read/write request.

TABLE 7: RANDOM REQUESTS TABLE WITH FIXED NUMBER OF REQUEST (1000)

| The number of request=1000 | | |
| --- | --- | --- |
| Cost of *ORAD* algorithm | Cost of *ADRW* algorithm | Mean probability of read request |
| 16409 | 16271 | 0.1 |
| 16177 | 16364 | 0.2 |
| 15590 | 15824 | 0.3 |
| 15548 | 15727 | 0.4 |
| 16188 | 16466 | 0.5 |
| 15964 | 16560 | 0.6 |
| 15392 | 16856 | 0.7 |
| 13127 | 13758 | 0.8 |
| 10542 | 10124 | 0.9 |

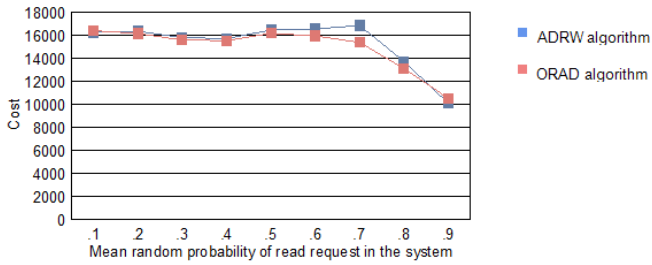

Fig. 8 Cost performance of *ORAD* and *ADRW* algorithm when the number of request is fixed (1000) and each node has the same probability of read/write request

As Fig. 6, it is clear in Fig. 7 and Fig. 8 that *ORAD* algorithm can perform better than *ADRW* algorithm. Because the number of requests in each probability of read request is the same, we can conclude that these experiments (Fig. 7 and Fig. 8) are more useful than Fig. 6 to compare these tow algorithms. We also observe that the average cost of servicing a request improved about 2.34 percent in Fig. 7 and 2.18 percent in Fig. 8 by *ORAD* algorithm compared with *ADRW* algorithm.

Further, we also perused the performance of these tow algorithms with several request sequences as the Table 8.

In Fig. 9, we observe the cost performance of these tow algorithms in several request sequences. Fig. 9 shows that the average cost of servicing a request improved about 5.68 percent by *ORAD* algorithm in comparison with *ADRW* algorithm.

Table 8: Request Sequence table

| Sequence name | Request sequence | Cost of *ORAD* algorithm | Cost of *ADRW* algorithm |
| --- | --- | --- | --- |
| $A$ | $R^{P3}_{O4}$, $R^{P6}_{O4}$, $W^{P2}_{O4}$, $R^{P3}_{O2}$, $W^{P7}_{O5}$, $R^{P4}_{O3}$, $W^{P5}$, $W^{P2}_{O3}$, $R^{P4}_{O2}$, $R^{P5}_{O1}$, $W^{P4}_{O5}$, $W^{P5}_{O1}$, $W^{P7}_{O5}$, $W^{P5}_{O1}$, $W^{P4}_{O1}$, $W^{P6}_{O4}$, $W^{P2}_{O2}$, $R^{P4}_{O4}$, $R^{P5}_{O3}$, $R^{P4}_{O5}$, $W^{P3}_{O4}$, $R^{P5}_{O2}$ | 429 | 455 |
| $B$ | $R^{P3}_{O2}$, $W^{P6}_{O4}$, $R^{P3}_{O4}$, $W^{P2}_{O4}$, $R^{P5}_{O4}$, $W^{P1}_{O5}$, $W^{P3}_{O2}$, $R^{P6}_{O2}$, $R^{P5}_{O3}$, $R^{P4}_{O3}$, $W^{P3}_{O2}$ | 180 | 188 |
| $C$ | $W^{P5}_{O4}$, $R^{P4}_{O3}$, $R^{P2}_{O4}$, $W^{P1}_{O5}$, $W^{P3}_{O5}$, $R^{P2}_{O4}$, $W^{P2}_{O1}$, $R^{P4}_{O3}$, $R^{P7}_{O5}$, $W^{P1}_{O1}$, $R^{P4}_{O5}$, $R^{P6}_{O2}$, $R^{P5}_{O2}$, $R^{P5}_{O1}$, $W^{P6}_{O2}$, $W^{P7}_{O5}$, $R^{P3}_{O4}$, $R^{P4}_{O5}$, $R^{P3}_{O2}$, $W^{P1}_{O5}$ | 303 | 323 |
| $D$ | $R^{P3}_{O2}$, $R^{P2}_{O2}$, $W^{P5}_{O2}$, $R^{P2}_{O3}$, $W^{P2}_{O3}$, $R^{P6}_{O2}$, $W^{P5}_{O2}$, $R^{P3}_{O2}$, $R^{P2}_{O1}$, $W^{P4}_{O3}$, $R^{P4}_{O3}$, $R^{P3}_{O2}$ | 246 | 253 |
| $E$ | $R^{P4}_{O3}$, $R^{P6}_{O4}$, $W^{P6}_{O2}$, $R^{P5}_{O3}$, $W^{P7}_{O3}$, $W^{P4}_{O5}$, $W^{P2}_{O2}$, $R^{P5}_{O3}$, $W^{P2}_{O1}$, $R^{P4}_{O3}$, $R^{P6}_{O4}$, $W^{P6}_{O2}$, $R^{P2}_{O2}$, $R^{P5}_{O2}$, $R^{P7}_{O5}$ | 242 | 267 |
| $F$ | $R^{P3}_{O2}$, $R^{P2}_{O2}$, $W^{P4}_{O5}$, $R^{P6}_{O4}$, $W^{P2}_{O2}$, $R^{P2}_{O3}$, $R^{P1}_{O5}$, $R^{P4}_{O4}$, $W^{P3}_{O4}$, $R^{P5}_{O4}$, $R^{P2}_{O1}$ | 194 | 204 |

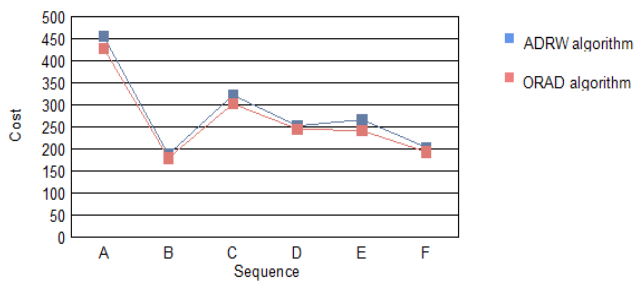

Fig. 9 Cost performance of *ORAD* and *ADRW* algorithm in several request sequences and each node has the same probability of read/write request

## 5. Conclusions

In this paper, we have proposed an optimum object replication algorithm, referred to as *ORAD* algorithm, for servicing random requests in distributed database systems. We explained the mechanism of *ORAD* algorithm with pictures. We also presented *ORAD* algorithm with using *TED/ TXD/ TF* policy. Our objective is to adjust the replica allocation that minimizes the access time over all servers and objects [7]. We simulated *ORAD* and *ADRW* algorithm on a PC based network and compared their performance under several conditions. We observed in the figures how each algorithm works in verify probability of read request and also, in several request sequences. In all experiments we saw that *ORAD* algorithm is superior to *ADRW* algorithm in the field of average request servicing cost and it is because of two used tricks in the mechanism of *ORAD* algorithm[1].

From the above experiments, we can conclude that if the mean probability of read request in a system is certain or uncertain, it is recommended to use *ORAD* algorithm. It is because *ORAD* algorithm can obtain the minimum average cost for servicing a request.

**Arash Ghorbannia Delavar** received the MSc and Ph.D. degrees in computer engineering from Sciences and Research University, Tehran, IRAN, in 2002 and 2007. He obtained the top student award in Ph.D. course. He is currently an assistant professor in the Department of Computer Science, Payam Noor University, Tehran, IRAN. He is also the Director of Virtual University and Multimedia Training Department of Payam Noor University in IRAN. Dr.Arash Ghorbannia Delavar is currently editor of many computer science journals in IRAN. His research interests are in the areas of computer networks, microprocessors, data mining, Information Technology, and E-Learning.

**Golnoosh Keshani** received a B.Sc. in computer engineering from Azad University, Mobarakeh, IRAN, in 2007, and M.Sc. student in computer engineering in Payam Noor University, Tehran, IRAN. Her research interests are in the areas of distributed databases, designing Algorithms, analyzing systems, designing databases.

---

[1] In the mechanism of *ORAD* algorithm, a data processor of an object can write on the object locally and it is not required that the server sends the object to it. Furthermore, in this mechanism, a non-data processor of an object can read temporary the object from its local memory.